\documentclass[conference,9pt]{IEEEtran}

\usepackage[english]{babel}
\usepackage[utf8]{inputenc}

\usepackage[pdftex]{graphicx}
\usepackage{epstopdf}

\usepackage{float}
\usepackage{caption}

\usepackage{amsmath}           

\usepackage{amsthm}
\usepackage{amsfonts}

\begin{document}

\title{Full-Duplex Relaying in MIMO-OFDM Frequency-Selective Channels with Optimal Adaptive Filtering}

\author{
    \IEEEauthorblockN{João S. Lemos\IEEEauthorrefmark{1}, 
     Francisco A. Monteiro\IEEEauthorrefmark{2}, 
	 Ivo Sousa\IEEEauthorrefmark{1},    
     and António Rodrigues\IEEEauthorrefmark{1}}
    \IEEEauthorblockA{\IEEEauthorrefmark{1}Instituto de Telecomunicações, and Instituto Superior Técnico, Universidade de Lisboa, Portugal}
    \IEEEauthorblockA{\IEEEauthorrefmark{2}Instituto de Telecomunicações, and ISCTE - Instituto Universitário de Lisboa, Portugal}
   
email: joao.sande.lemos@tecnico.ulisboa.pt , francisco.monteiro@lx.it.pt

}
\maketitle

\begin{abstract}

In-band full-duplex transmission allows a relay station to theoretically double its spectral efficiency by simultaneously receiving and transmitting in the same frequency band, when compared to the traditional half-duplex or out-of-band full-duplex counterpart. Consequently, the induced self-interference suffered by the relay may reach considerable power levels, which decreases the signal-to-interference-plus-noise ratio (SINR) in a decode-and-forward (DF) relay, leading to a degradation of the relay performance. This paper presents a technique to cope with the problem of self-interference in broadband multiple-input multiple-output (MIMO) relays. The proposed method uses a time-domain cancellation in a DF relay, where a replica of the interfering signal is created with the help of a recursive least squares (RLS) algorithm that estimates the interference frequency-selective channel. Its convergence mean time is shown to be negligible by simulation results, when compared to the length of a typical orthogonal-frequency division multiplexing (OFDM) sequences. Moreover, the bit-error-rate (BER) and the SINR in a OFDM transmission are evaluated, confirming that the proposed method extends significantly the range of self-interference power to which the relay is resilient to, when compared with other mitigation schemes.                 
\linebreak

 
\end{abstract}

\section{Introduction}
\label{sec:INTRO}

A relay station is a key element in a wireless multi-hop network, which is believed to incorporate future communication systems, since it may provide a wider coverage, a higher data rate and a lower transmit peak power. Therefore, a considerable amount of research has been conducted in this topic, where the integration of multiple-input multiple-output (MIMO) techniques at the relay appear as a natural solution to avoid the key-hole effect \cite{almers:keyhole}.

In-band full-duplex operation is a novel technique, which has recently gained attention in the wireless communication field \cite{ibfd:challenges}. Its main advantage is the possibility to double the spectral efficiency of the relay station. This is achieved by employing the same time and frequency resources for receiving and transmitting, i.e., the relay  operation of receive and forward information is performed within the same frequency band at the same time. However, the clear limitation of full-duplex operation is the arising loopback self-interference due to the leakage of the relay outgoing signal to the relay receiver side, mainly enhanced by the high power unbalance between the desired and the self-interference signal, causing inadmissible levels of interference and deteriorating the relay performance \cite{melissa:ffr}. Thus, the self-interference is typically attenuated by a proper system physical design and by subtracting a delayed version of the own signal at an analog stage \cite{jain:practical}. Then, a digital baseband processing stage is often introduced \cite{riih:mit}, \cite{himal:fdeteperformance}, \cite{hien:multipair}, so that a reliable communication link may be established. 

The relaying protocol often employed in full-duplex systems is characterized by regenerating the message from the original source, known as decode-and-forward (DF) operation. When compared to amplify-and-forward (AF), that blindly repeats the ongoing message, DF additional complexity grants an improvement in terms of performance \cite{nos:coop}. Among these digital DF techniques stand two methods, time-domain cancellation, subtracting a baseband version of the self-interference signal, and interference suppression, which exploits the degrees of freedom available in MIMO channels \cite{riihonen:rateAnalysis,riih:mit,lemos:MMrelay}. The first method chiefly suffers from self-interference erroneous channel estimation, while the latter is affected by the distortion from spatial shaping of the relay transmitted signal. Besides the aforementioned, both performances are deteriorated by limited dynamic range and transmission impairments that introduce additional noise sources \cite{emilio:sinropt}. Those effects can drastically harm the operation of full-duplex relaying, boosting the search for better mitigation schemes. 

Adaptive cancellation was firstly proposed in \cite{emilio:adapt}, where the authors make use of a gradient-descent algorithm to estimate the self-interference and achieve a 28.6 dB of additional interference attenuation. This paper proposes for the first time, to the best of our knowledge, the use of recursive least squares (RLS) filtering to further improve that resilience to self-interference, which converges to the optimal mean-square error (MSE) estimator in negligible time.   
The well-known RLS algorithm \cite{haykin:adapfilter} is adapted to the MIMO full-duplex relay. Moreover, it is a suitable technique to cope with frequency-selective effects present in broadband MIMO transmissions, by exploiting orthogonal-frequency division multiplexing (OFDM) transmissions \cite{mont:FSmimo}.            
Moreover, this scheme acts as an add-on block to the half-duplex relay, i.e., it does not change the relay protocol and can be added whenever the full-duplex mode is enabled.       

\section{System Model}
\label{sec:SM}

\begin{figure*}
\begin{center}
\includegraphics[trim=110 210 110 180, scale=0.55]{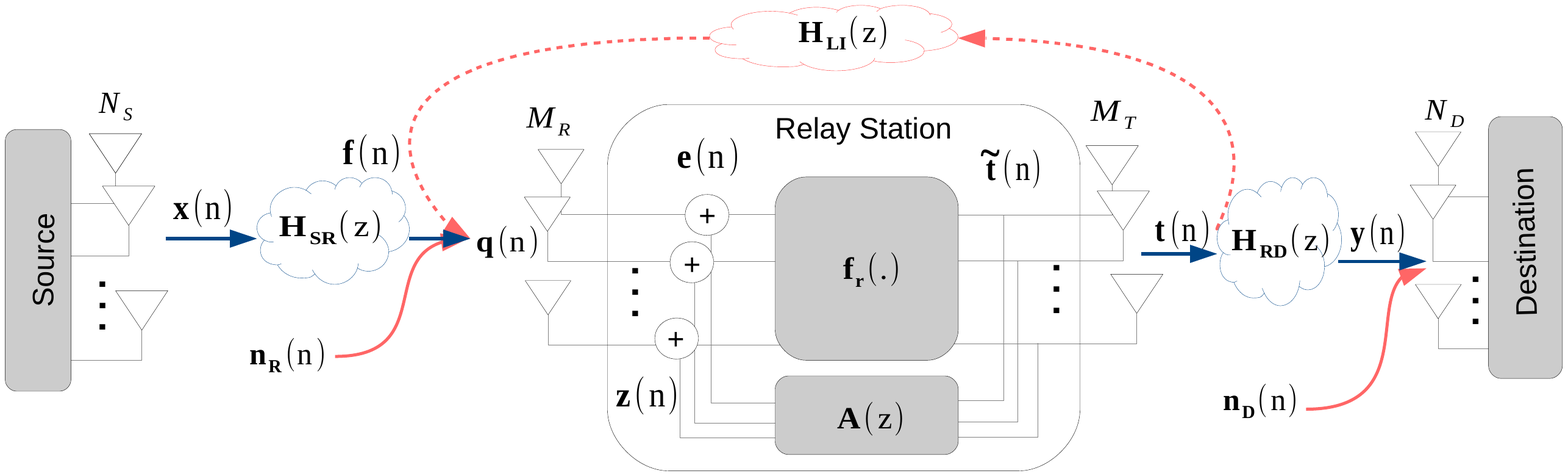}
\end{center}
\caption{Model of a decode-and-forward in-band full-duplex relay station with adaptive self-interference cancellation.}
\label{fig:relay_feedback}
\end{figure*}

Consider a single-hop wireless communication through a MIMO relay station, which operates in a decode-and-forward full-duplex mode. The relay serves a source with a total of $N_S$ antennas that intends to establish a communication link to a destination with $N_D$ antennas. The relay is equipped with $M_R$ receiving antennas and $M_T$ transmitting antennas. Figure \ref{fig:relay_feedback} depicts the considered broadband uncoded MIMO transmission through a full-duplex relay with self-interference cancellation.      

At time instant $n$, a source signal $\mathbf{x}(n) \in \mathbb{C}^{N_S}$ is transmitted via a relay station to the destination. The relay station receives a version of that signal, $\mathbf{q}(n) \in \mathbb{C}^{M_R}$, after passing through a frequency-selective channel and being mixed with the self-interference and additive noise components. At the same time, the relay regenerates and retransmits a signal, $\mathbf{t}(n) \in \mathbb{C}^{M_T}$, which reaches the destination as $\mathbf{y}(n) \in \mathbb{C}^{N_D}$, also experiencing channel and noise effects. Equation \eqref{eq:FDrelayreceivedsignal} shows the expressions for both $\mathbf{q}(n)$ and $\mathbf{y}(n)$
\begin{equation}
\begin{split}
    \label{eq:FDrelayreceivedsignal}
    \mathbf{q}(n)=&\mathbf{H_{SR}}(z)\mathbf{x}(n)+\mathbf{H_{LI}}(z)\mathbf{t}(n)+\mathbf{n_R}(n),\\
    \mathbf{y}(n)=&\mathbf{H_{RD}}(z)\mathbf{t}(n)+\mathbf{n_D}(n),
\end{split}
\end{equation}   
where $\mathbf{H_{SR}}(z) \in \mathbb{C}^{M_R \times N_S}$ is the $L_{SR}$ order channel from the source to the relay, $\mathbf{H_{RD}}(z) \in \mathbb{C}^{N_D \times M_T}$ is the $L_{RD}$ order channel from the relay to the destination, and ${\mathbf{H_{LI}}(z) \in \mathbb{C}^{M_R \times M_T}}$ is the $L_{LI}$ order self-interference channel. Both frequency-selective channels matrices are represented by their $z$-transform, ${\mathbf{H}(z)=\sum_{k=0}^{L}\mathbf{H}[k]z^{-k}}$, where $z$ represents the discrete delay operator and $L$ the filter order; e.g., $\mathbf{H_{SR}}(z)\mathbf{x}(n)=\sum_{k=0}^{L_{SR}}\mathbf{H_{SR}}[k]\mathbf{x}(n-k)$. The vectors $\mathbf{n_R}(n)$ and $\mathbf{n_D}(n)$ are the additive noise components at the input of the relay station and destination, respectively. 

The relay DF protocol $f_r(\cdot)$ is independent of the self-interference cancellation architecture, and regenerates a delayed estimation of the received data streams, $\mathbf{t}(n)=f_r(\mathbf{e}(n-d),\mathbf{e}(n-d-1),\cdots,\mathbf{e}(n-d-D))$, where $d$ stands for the necessary processing delay and $D+1$ is the length of the employed time-window. Further, the processing delay is considered strictly positive and sufficiently long \cite{riih:mit}, such that $\mathbf{x}(n-k)$ and $\mathbf{t}(n-l)$ become uncorrelated, i.e., $\mathbb{E}\{\mathbf{x}(n-k)\mathbf{t}^H(n-l)\}=\mathbf{0}$, for all $k=0,\cdots,L_{SR}$ and all $l=0,\cdots,L_{LI}$. 

The practical implementation impairments in the relay, arising from imperfect AD/DA conversion, limited dynamic range, quantization errors, power amplifier non-linearities, etc., are jointly modeled as additive interference noise \cite{emilio:sinropt}, giving ${\mathbf{{t}}(n)=\mathbf{\tilde{t}}(n)+\mathbf{\mathcal{E}_{t}}(n)}$, where $\mathbf{\tilde{t}}(n)$ is the known baseband signal at the relay. The error $\mathbf{\mathcal{E}_{t}}(n)$, takes into consideration the relay transmit power weighted by factor $\delta$, and is characterized by
\begin{equation}
\begin{split}
\label{eq:impair}
\mathbf{\mathcal{E}_{t}}(n) \sim \mathcal{CN}\big(\mathbf{0}, (\delta \cdot \mathbb{E} \{\mathbf{\tilde{t}}^H(n)\mathbf{\tilde{t}}(n) \} ) \mathbf{I}\big).
\end{split}
\end{equation}  
Moreover, the relay may use an estimation of the self-interference channel matrix, $\mathbf{\tilde{H}_{LI}}(z)$, when necessary. However, this estimation is associated with an additive noise term, $\mathbf{\mathcal{E}_{{H}_{LI}}}[k](n)$, which models the practical non-ideal channel estimation in such relay systems   
\begin{equation}
\begin{split}
    \label{eq:FDmatrixerror}
    \mathbf{{H}_{LI}}[k]&=\mathbf{\tilde{H}_{LI}}[k]+\mathbf{\mathcal{E}_{{H}_{LI}}}[k](n),\text{ for }k=0,\cdots,L_{LI}.\\
\end{split}
\end{equation}   


\section{Mitigation of Self-Interference}
\label{sec:MSI}

In order to properly mitigate the self-interference present in the relay pre-decoding signal $\mathbf{e}(n)$, we apply a feedback canceler filter in the time domain. The idea is to estimate the self-interference channel effect on the transmitted signal, given by $\mathbf{f}(n)=\mathbf{H_{LI}}(z)\mathbf{{t}}(n)$, employing a finite impulse response (FIR) filter $\mathbf{A}(z) \in \mathbb{C}^{M_R \times M_T}$, with order $L_A$, that simulates the self-interference chain effect, from the relay transmit side to the receive one. Therefore, we have
\begin{equation}
\begin{split}
    \label{eq:FDprefunction}
    \mathbf{e}(n)&=\mathbf{q}(n)+\mathbf{{z}}(n)\\
    	&=\mathbf{H_{SR}}(z)\mathbf{x}(n)+\mathbf{H_{LI}}(z)\mathbf{t}(n)+\mathbf{n_R}(n)+\mathbf{A}(z)\mathbf{\tilde{t}}(n).
\end{split}
\end{equation} 


\subsection{Time-Domain Cancellation}
\label{tdc}

Time-domain cancellation (TDC) is the trivial solution to the problem presented above, as it uses the self-interference channel estimation to perform the cancellation \cite{riih:mit}, as 
\begin{equation}
\begin{split}
    \label{eq:FDfiltertdc}
    \mathbf{A}_{TDC}(z)=-\mathbf{\tilde{H}_{LI}}(z),
\end{split}
\end{equation} 
where the filter order should be $L_A = L_{LI}$. However, this procedure is severely harmed by the above mentioned system impairments, as we cannot remove the following residual interference 
\begin{equation*}
\begin{split}
    \label{eq:FDfiltersuberror}
    \mathbf{f}(n)+\mathbf{z}(n)&=\mathbf{H_{LI}}(z)\mathbf{t}(n)+\mathbf{A}_{TDC}(z)\mathbf{\tilde{t}}(n)\\
    &=  \mathbf{\mathcal{E}_{H_{LI}}}(z)\Big(\mathbf{\tilde{t}}(n)+\mathbf{\mathcal{E}_{t}}(n)\Big) +  \mathbf{\tilde{H}_{LI}}(z)\mathbf{\mathcal{E}_{t}}(n),
\end{split}
\end{equation*}
being this technique performance dependent on the power of the above residual terms.

\subsection{Recursive Least Squares Cancellation}
The proposed cancellation scheme is based only on the transmitted signal $\mathbf{\tilde{t}}(n)$ and observed signal $\mathbf{q}(n)$, easily accessed by the relay, to estimate the self-interference effect. We start by considering the mean square error of the self-interference channel estimation effect at time instant $n$, $ MSE(n,\mathbf{A}(z))=\mathbb{E}\{(\mathbf{{f}}(n) -   \mathbf{A}(z)\mathbf{\tilde{t}}(n))^H (\mathbf{{f}}(n) -  \mathbf{A}(z)\mathbf{\tilde{t}}(n) )\} $, for the generalized MIMO case, and defining the recursive least squares approximation of it as
\begin{equation}
\begin{split}
    \label{eq:mserls}
	& \widehat{MSE}_{RLS}(n,\mathbf{A}(z)) =\\ 
	&=\sum_{k=1}^n \lambda^{n-k} \big(\mathbf{{f}}(k) -   \mathbf{A}(z)\mathbf{\tilde{t}}(k)\big)^H 
	 \big(\mathbf{{f}}(k) -   \mathbf{A}(z)\mathbf{\tilde{t}}(k)\big),
\end{split}       
\end{equation}
where $0 < \lambda \leq 1$ is the forgetting factor, which regulates the dependency on previous observations. The RLS algorithm is derived by exploiting the second derivative of the $\widehat{MSE}_{RLS}$, as a Newton-type algorithm \cite{haykin:adapfilter}. Defining the concatenation matrix of the filter parameters $\mathbf{A}_{\star}(n) = \big[\mathbf{A}[0](n),\cdots,\mathbf{A}[L_A](n)\big]^T \in\ \mathbb{C}^{  M_T(L_A+1) \times M_R}$, the RLS correlation matrices as 
$$\mathbf{\hat{\Sigma}}_{\mathbf{\tilde{t}},\mathbf{{f}}}^{RLS}(n)=\sum_{k=1}^n \lambda^{n-k} \mathbf{{f}}^H(k)\otimes\mathbf{\bar{t}}(k), \in \mathbb{C}^{ {(L_A+1)} M_T \times M_R},$$ 
$$\mathbf{\hat{\Sigma}}_{\mathbf{\tilde{t}},\mathbf{\tilde{t}}}^{RLS}(n)=\sum_{k=1}^n \lambda^{n-k} \mathbf{R_{\mathbf{\tilde{t}}}}(k), \in \mathbb{C}^{ ({L_A}+1) M_T \times (L_A+1) M_T},$$ 
where $ \mathbf{R_{\mathbf{\tilde{t}}}}(k) = \mathbf{\bar{t}}(n)\mathbf{\bar{t}}^H(n)$, with $ \mathbf{\bar{t}}(n) = \text{vec}\{\mathbf{\tilde{T}}(n)\} = \text{vec}\{\big[\mathbf{\tilde{t}}(n),\cdots,\mathbf{\tilde{t}}(n-L_A) \big]\} \in \mathbb{C}^{ ({L_A}+1) M_T \times 1}$ is a column vector containing the elements of $\mathbf{\tilde{T}}(n)$, and $\otimes$ represents the Kronecker product,
it is possible to write the Newton algorithm time update rule by taking the first and second derivatives of equation \eqref{eq:mserls}, as 
\begin{equation}
\begin{alignedat}{2}
    \label{eq:RLS1}
    &\mathbf{\hat{A}}_{\star, n} =\mathbf{\hat{A}}_{\star, n-1} -\frac{\mu}{2}\Big[\frac{\partial^2}{\partial\mathbf{A}_{\star}^2} \widehat{MSE}_{RLS}(n,\mathbf{\hat{A}}_{\star}(n-1)) \Big]^{-1}\cdot\\
    &\qquad  \frac{\partial}{\partial\mathbf{{A}}_{\star}} \widehat{MSE}_{RLS}(n,\mathbf{\hat{A}}_{\star}(n-1))  \\
    &=\mathbf{\hat{A}}_{\star, n-1} + \mu \big[\mathbf{\hat{\Sigma}}_{\mathbf{\tilde{t}},\mathbf{\tilde{t}}}^{RLS}(n) \big]^{-1} \Big( \mathbf{\hat{\Sigma}}_{\mathbf{\tilde{t}},\mathbf{{f}}}^{RLS}(n) - 
   \mathbf{\hat{\Sigma}}_{\mathbf{\tilde{t}},\mathbf{\tilde{t}}}^{RLS}(n) \mathbf{\hat{A}}_{\star, n-1} \Big),
\end{alignedat}
\end{equation}
where $\mu$ is the Newton-type algorithm step size.

However, equation \eqref{eq:RLS1} still depends on $\mathbf{f}(n)$, which is in fact not available at the relay station. Further, the algorithm requires at each iteration the inversion of matrix $\mathbf{\hat{\Sigma}}_{\mathbf{\tilde{t}},\mathbf{\tilde{t}}}^{RLS}(n)$, costing $ \mathcal{O} \big( ( (L_A+1) M_T )^3 \big)$.   

\section{Algorithm Analysis}
\label{sec:AA}
In this section we provide an efficient update rule for the proposed algorithm, using the available relay input signal, $\mathbf{q}(n)$. Also, the asymptotic error covariance matrix is evaluated.

\subsection{Update Rule}
\label{UP}
Firstly, we notice that the correlation between $\mathbf{q}(n)$ and $\mathbf{\tilde{t}}(n)$, assuming weakly stationary variables, is given by
\begin{equation}
\begin{split}
    \label{eq:FDRqt}
    \mathbf{R}_{\mathbf{q},\mathbf{\mathbf{\tilde{t}}}}(k)\ &=\ \mathbb{E}\{\mathbf{q}(n) \mathbf{\tilde{t}}^H(n-k)\}\\
    &=\ \mathbb{E}\{(\mathbf{H_{SR}}(z)\mathbf{x}(n)+\mathbf{f}(n)+\mathbf{n_R}(n))\mathbf{\tilde{t}}^H(n-k) \}\\
    &=\ \mathbb{E}\{ \mathbf{f}(n) \mathbf{\tilde{t}}^H(n-k)\}\},
\end{split}
\end{equation} 
where $\mathbb{E}\{ \mathbf{x}(n) \mathbf{\tilde{t}}^H(n-k) \}=\mathbf{0}$, for a sufficiently large processing delay, and also $\mathbb{E}\{ \mathbf{n_R}(n)\mathbf{\tilde{t}}^H(n-k) \}=\mathbf{0}$. This result allows to observe $\mathbf{q}(n)$ instead of $\mathbf{f}(n)$, since the correlation matrix in \eqref{eq:FDRqt} is $\mathbf{R}_{\mathbf{{q}},\mathbf{\tilde{t}}}(k)=\mathbf{R}_{\mathbf{{f}},\mathbf{\tilde{t}}}(k)$, therefore, not changing the filter convergence properties. Nevertheless, the equivalent noise present in the observed signal $\mathbf{q}(n)$ is now composed
by the source and additive noise vectors, which will affect the filter performance, mainly its convergence time.

Resorting to the Woodbury matrix identity \cite{haykin:adapfilter}, by defining $\mathbf{\bar{P}}(n)=\big[\mathbf{\hat{\Sigma}}_{\mathbf{\tilde{t}},\mathbf{\tilde{t}}}^{RLS}(n) \big]^{-1}$, it is possible to show that  
\begin{equation}
\begin{split}
    \label{eq:RLS2}
  	\mathbf{\bar{P}}(n+1)=&\big[\lambda\mathbf{\hat{\Sigma}}_{\mathbf{\tilde{t}},\mathbf{\tilde{t}}}^{RLS}(n-1)+\mathbf{R_{\mathbf{\tilde{t}}}}(n)\big]^{-1}\\
  	=&\frac{1}{\lambda}\big[\mathbf{\bar{P}}^{-1}(n)+\frac{1}{\lambda} \mathbf{R_{\mathbf{\tilde{t}}}}(n)\big]^{-1}\\
  	=&\frac{1}{\lambda}\big[\mathbf{\bar{P}}^{-1}(n)+ \mathbf{\bar{t}}(n)\frac{1}{\lambda}\mathbf{\bar{t}}^H(n)\big]^{-1}\\
  	=&\frac{1}{\lambda}\Big( \mathbf{\bar{P}}(n)- \frac{\mathbf{\bar{P}}(n)\mathbf{R_{\mathbf{\tilde{t}}}}(n)\mathbf{\bar{P}}(n)}{\lambda+\mathbf{\bar{t}}^H(n)\mathbf{\bar{P}}(n)\mathbf{\bar{t}}(n)} \Big),
\end{split}
\end{equation}
\looseness -1 which does not require the matrix inversion in \eqref{eq:RLS1}, reducing the update rule complexity order from cubic to quadratic. 

Thus, we can finally present the algorithm 3-step update rule as
\begin{equation}
\begin{split}
    \label{eq:RLS2}
  	 &\mathbf{k}(n)\ =\ \frac{\mathbf{\bar{P}}(n)\mathbf{\bar{t}}(n)}{\lambda+\mathbf{\bar{t}}^H(n)\mathbf{\bar{P}}(n)\mathbf{\bar{t}}(n)}\ ,\\
  	 &\mathbf{\hat{A}}_{\star, n}=\mathbf{\hat{A}}_{\star, n-1} + \mu \mathbf{k}(n)\big(\mathbf{q}(n) -  \sum_{l=0}^{L_A}\mathbf{\hat{A}}[l](n-1)\mathbf{\tilde{t}}(n-l)  \big)^H,\\
  	 &\mathbf{\bar{P}}(n+1)\ =\  \frac{1}{\lambda}\Big(\mathbf{\bar{P}}(n+1) - \mathbf{k}(n)\mathbf{\bar{t}}^H(n)\mathbf{\bar{P}}(n)\Big),  
\end{split}
\end{equation}   
where vector $\mathbf{k}(n)$ is the update direction of the filter \cite{haykin:adapfilter}.

\subsection{Asymptotic Error Covariance}
In this subsection, we prove that the algorithm converges to the optimal value by evaluating the asymptotic error covariance matrix of the iterative expressions in \eqref{eq:RLS2}. For that end, we assume the case of no forgetting factor ($\lambda=1$), which guarantees the filter convergence \cite{ele:rls}.
Considering the Newton update rule in \eqref{eq:RLS1} and the observation made in \ref{UP}, the filter parameters at each iteration may be given by $\mathbf{\hat{A}}_{\star, n}=\big[\mathbf{\hat{\Sigma}}_{\mathbf{\tilde{t}},\mathbf{\tilde{t}}}^{RLS}(n) \big]^{-1} \mathbf{\hat{\Sigma}}_{\mathbf{\tilde{t}},\mathbf{{q}}}^{RLS}(n) = \big[\frac{1}{n}\mathbf{\hat{\Sigma}}_{\mathbf{\tilde{t}},\mathbf{\tilde{t}}}^{RLS}(n) \big]^{-1} \frac{1}{n}\mathbf{\hat{\Sigma}}_{\mathbf{\tilde{t}},\mathbf{{q}}}^{RLS}(n)$. Thus, as $n$ grows large, the RLS covariance matrices will converge to the true covariance matrices, $\mathbf{{\Sigma}}_{\mathbf{\tilde{t}},\mathbf{\tilde{t}}}^{RLS}(n)\rightarrow\mathbf{{\Sigma}}_{\mathbf{\tilde{t}},\mathbf{\tilde{t}}}=\mathbb{E}\{ \mathbf{\bar{t}}(n)\mathbf{\bar{t}}^H(n) \}$ and $\mathbf{{\Sigma}}_{\mathbf{\tilde{t}},\mathbf{{q}}}^{RLS}(n)\rightarrow\mathbf{{\Sigma}}_{\mathbf{\tilde{t}},\mathbf{{q}}}=\mathbb{E}\{ \mathbf{{q}}^H(n)\otimes\mathbf{\bar{t}}(k)\}$, by the law of large numbers \cite{ele:rls}. We may then write the error matrix at instant $n$ as in \eqref{eq:ErrorRLS} \cite{haykin:adapfilter}. 
\begin{equation}
\begin{split}
    \label{eq:ErrorRLS}
  	\mathbf{\tilde{A}}_{\star, n} =&\ \mathbf{\hat{A}}_{\star, n} -\mathbf{A}_{\star, Opt.}\\
    =&\ \big[\mathbf{\hat{\Sigma}}_{\mathbf{\tilde{t}},\mathbf{\tilde{t}}}^{RLS}(n) \big]^{-1} \mathbf{{\Sigma}}_{\mathbf{\tilde{t}},\mathbf{{q}}}^{RLS}(n)-\mathbf{{\Sigma}}_{\mathbf{\tilde{t}},\mathbf{\tilde{t}}} ^{-1} \mathbf{{\Sigma}}_{\mathbf{\tilde{t}},\mathbf{{q}}}\\
    =&\ \big[\mathbf{\hat{\Sigma}}_{\mathbf{\tilde{t}},\mathbf{\tilde{t}}}^{RLS}(n) \big]^{-1}           \sum_{k=1}^n  \mathbf{{q}}^H(k)\otimes\mathbf{\bar{t}}(k)-\mathbf{A}_{\star, Opt.}   \\
    \approx &\ \frac{1}{n} \mathbf{{\Sigma}}_{\mathbf{\tilde{t}},\mathbf{\tilde{t}}}^{-1}           \sum_{k=1}^n  \mathbf{{\tilde{f}}}^H(k)\otimes\mathbf{\bar{t}}(k),
\end{split}
\end{equation}   
where the last step comes from the above approximation, some algebraic manipulation, and defining ${\mathbf{{\tilde{f}}}(k)=\mathbf{{{q}}}(k)-\mathbf{A}_{ Opt.}(z)\mathbf{{\tilde{t}}}(k)}$ as the estimation error vector. Finally, we reach the covariance error parameter given by
\begin{equation}
\begin{split}
    \label{eq:ErrorRLS2}
  	&\mathbf{Q}(n) =\ \mathbb{E} \{ \mathbf{\tilde{A}}_{\star, n}\mathbf{\tilde{A}}_{\star,n}^H\}\\
  	  	&= \  \mathbf{{\Sigma}}_{\mathbf{\tilde{t}},\mathbf{\tilde{t}}} ^{-1}  \mathbb{E} \Big\{ \frac{1}{n^2}\sum_{k=1}^n  \mathbf{\tilde{f}}^H(k)\otimes\mathbf{\bar{t}}(k) \Big(\sum_{l=1}^n \mathbf{\tilde{f}}^H(l)\otimes\mathbf{\bar{t}}(l) \Big)^H \Big\}  \mathbf{{\Sigma}}_{\mathbf{\tilde{t}},\mathbf{\tilde{t}}}^{-1}\\
  	&= \  \mathbf{{\Sigma}}_{\mathbf{\tilde{t}},\mathbf{\tilde{t}}} ^{-1}  \frac{1}{n^2}\mathbb{E} \Big\{ \sum_{k=1}^n\sum_{l=1}^n  \mathbf{\bar{t}}(k)\mathbf{\tilde{f}}^H(k) \mathbf{\tilde{f}}(l) \mathbf{\bar{t}}^H(l) \Big\}  \mathbf{{\Sigma}}_{\mathbf{\tilde{t}},\mathbf{\tilde{t}}}^{-1}  \\
  	&= \ \frac{1}{n} \mathbb{E}\Big\{ \sum_{k=1}^n\sum_{l=1}^n\mathbf{\tilde{f}}^H(k) \mathbf{\tilde{f}}(l)\Big\}  \mathbf{{\Sigma}}_{\mathbf{\tilde{t}},\mathbf{\tilde{t}}}^{-1},  
  	\end{split}
\end{equation}
where it was assumed that $\mathbf{\tilde{f}}(n)$ and $\mathbf{\bar{t}}(n)$ are mutually independent. Thus, the error power of the RLS estimation for $\lambda=1$ converges to zero as the number of iterations $n$ tend to infinity, which provides an optimal estimation of the self-interference matrix.

\section{Numerical Results}
\label{sec:NR}

In this section, we study the proposed RLS filter performance, in terms of its convergence time, SINR and bit-error-rate (BER), evaluated at the relay station. All results are obtained via Monte-Carlo simulation of 2000 OFDM transmitted symbols, for a typical relay station scenario. 

\subsection{System Parameters}

We consider an uncoded MIMO-OFDM system, where $N_S=2$ data streams transmit a 16-QAM modulated OFDM stream of symbols with a block of $N_{sub}$ subcarriers and a cyclic prefix of $N_{cp}=L_{SR}$, to a destination also with $N_D=2$ antennas. The relay is considered symmetric and composed by $M_R=3$ receive antennas and $M_T=3$ transmit antennas, with a proper rate adaptation. Further, the channels are assumed to be of order one, i.e., $L_{SR}=L_{RD}=L_{LI}=1$ (2 taps channels), similar to what is considered in \cite{emilio:adapt}. The channels $\mathbf{H_{SR}}(z)$ and $\mathbf{H_{RD}}(z)$ are drawn from complex Gaussian distributions and taken from $\mathcal{CN}(0,1)$, while each self-interference matrix channel tap has distribution ${\mathbf{H_{LI}}[k] \sim \mathcal{CN}(0,\sigma^2_{LI} \mathbf{I} )}$, for $k=0, \cdots, L_{LI}$, where $\sigma^2_{LI}$ accounts for the residual power of the self-interference channel, after propagation and analog-circuit first stage of mitigation. The filter order is set to be $L_A=L_{LI}=1$, so that we can achieve a perfect estimation of the self-interference channel.    
The transmitted source and relay signals have normalized power, i.e., before OFDM modulation we have $\mathbb{E}\{ \mathbf{x}^H(n) \mathbf{x}(n)\}=1$ and $\mathbb{E}\{ \mathbf{\tilde{t}}^H(n) \mathbf{\tilde{t}}(n)\}=1$. Additionally, the thermal noise present at the relay, $\mathbf{n_R}(n)$, is considered to be additive white Gaussian with distribution $\mathbf{n_R}(n)  \sim \mathcal{CN}(0,\sigma^2_{\mathbf{n_R}} \mathbf{I})$, where its power is set to ${\sigma^2_{\mathbf{n_R}}=-15 \text{ dB}}$. Finally, the filter update rule is as defined in equation \eqref{eq:RLS2} with no forgetting factor $(\lambda=1)$ and with step size $\mu=1$. 

\subsection{Convergence Time}

\begin{figure}[t]
\begin{center}
\includegraphics[ width=.85\columnwidth, trim=4mm 0.1mm 5mm 6.9mm, clip=true, draft=false]{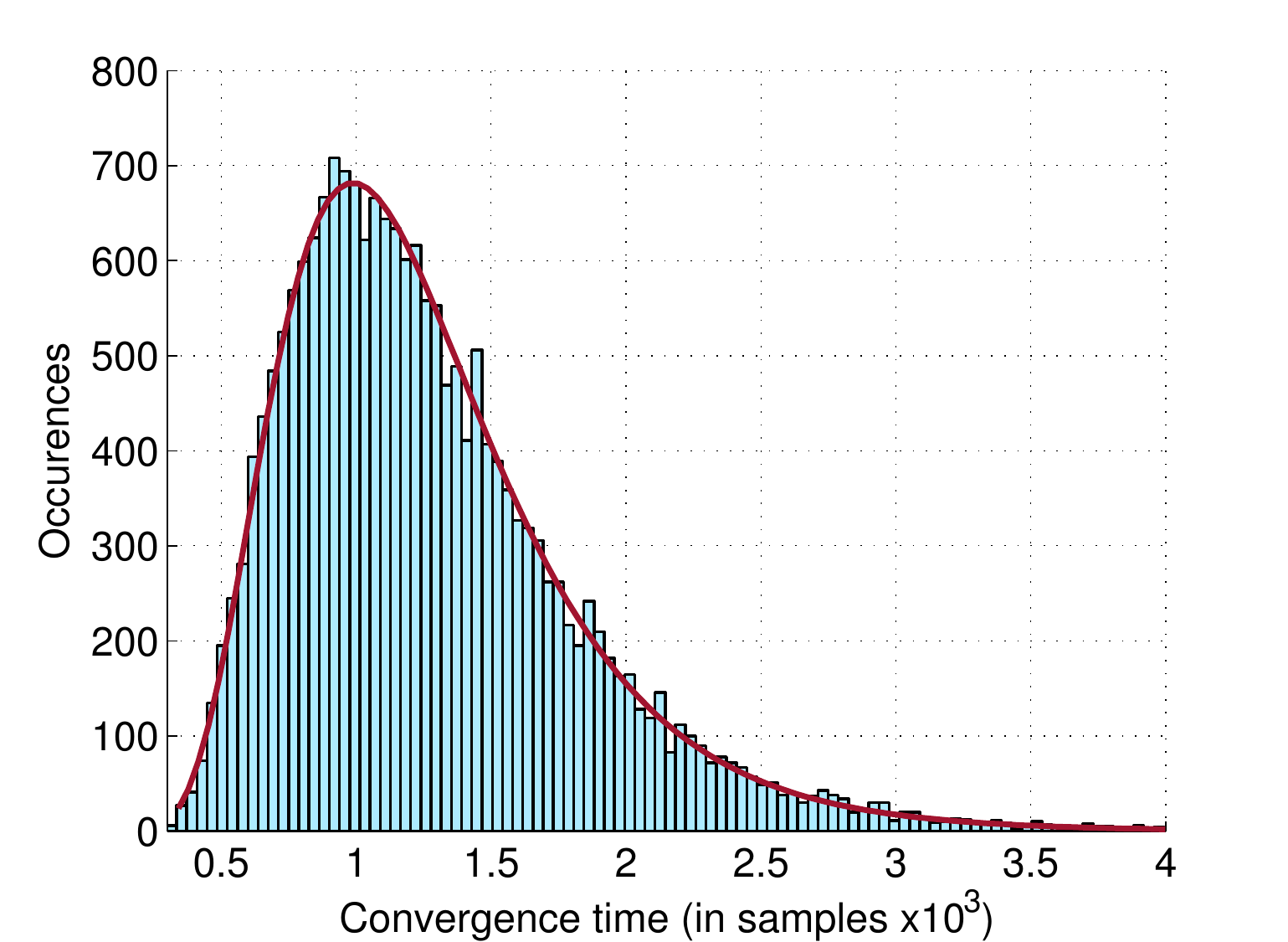}
\end{center}
\caption{Histogram of the proposed RLS algorithm convergence time for the self-interference channel estimation.}
\label{fig:hist}
\end{figure}

Let us firstly analyse the convergence time of the RLS algorithm, i.e., the number of iterations required from an initial starting point, $\mathbf{\hat{A}}_{\star, 0}=\mathbf{0}$ and $\mathbf{\bar{P}}(1)=\mathbf{I}$, to a certain point where the filter parameters satisfy the following error~metric~($EM$) \cite{emilio:adapt}
\begin{equation}
\begin{split}
    \label{eq:LMSerrorMS4}
   EM\leq\frac{\parallel\mathbf{\hat{A}}_{\star, n} - \mathbf{{H}_{LI,\star}} \parallel_F^2}{\parallel\mathbf{{H}_{LI,\star}}\parallel_F^2},
\end{split}
\end{equation}       
where $\mathbf{{H}_{LI,\star}}$ is the matrix concatenation of the self-interference channel taps, $\mathbf{{H}_{LI,\star}}=\big[\mathbf{{H}_{LI}}[0],\cdots,\mathbf{{H}_{LI}}[L_A]\big]^T$.

To that end, we consider an OFDM transmission with $N_{sub}=8192$ subcarriers and a self-interference channel with power $\sigma^2_{LI}= 0$ dB. The implementation impairments are taken into consideration by setting $\delta=10^{-5}$, as in \eqref{eq:impair}.  Figure~\ref{fig:hist} shows the histogram of the self-interference channel estimation RLS algorithm convergence time distribution, considering ${EM\leq-30}$ dB, after 20000 realizations. Further, a superimposed log-normal distribution indicates that the mean value of the algorithm convergence time is 1007 samples. This number is in fact less than the OFDM symbol length considered for simulation, and negligible among the number of transmitted OFDM sequences in real implementations.

\subsection{SINR and BER} 
 
\begin{figure}[t]
\begin{center}
\includegraphics[width=0.85\columnwidth, trim=4mm 0mm 5mm 1.8mm, clip=true, draft=false]{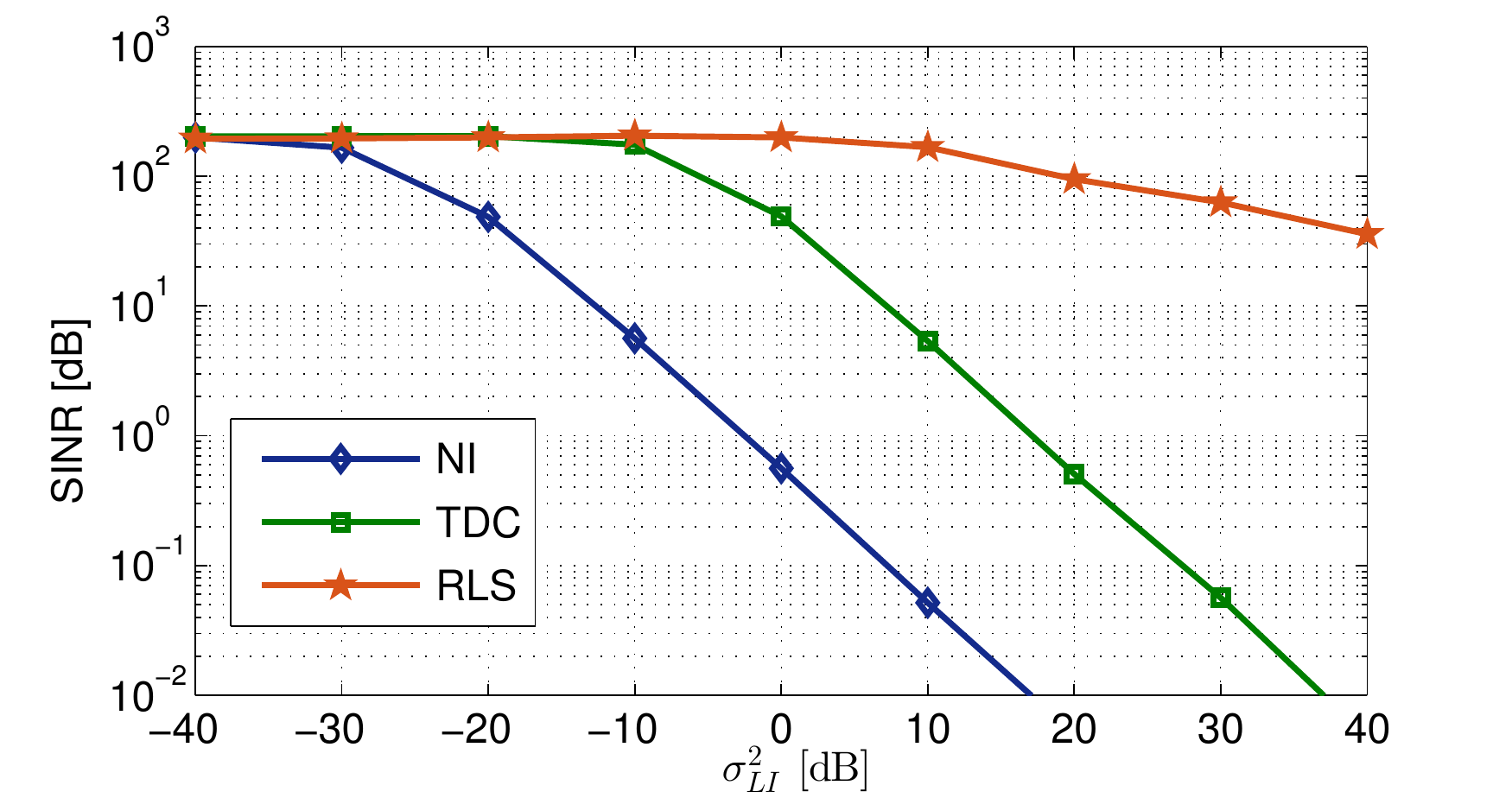}
\end{center}
\caption{Impact on the SINR of the self-interference power for the RLS, TDC and NI methods.}
\label{fig:SINR}
\end{figure}
\begin{figure}[t]
\begin{center}
\includegraphics[width=0.85\columnwidth, trim=2mm 0mm 5mm 3.5mm, clip=true, draft=false]{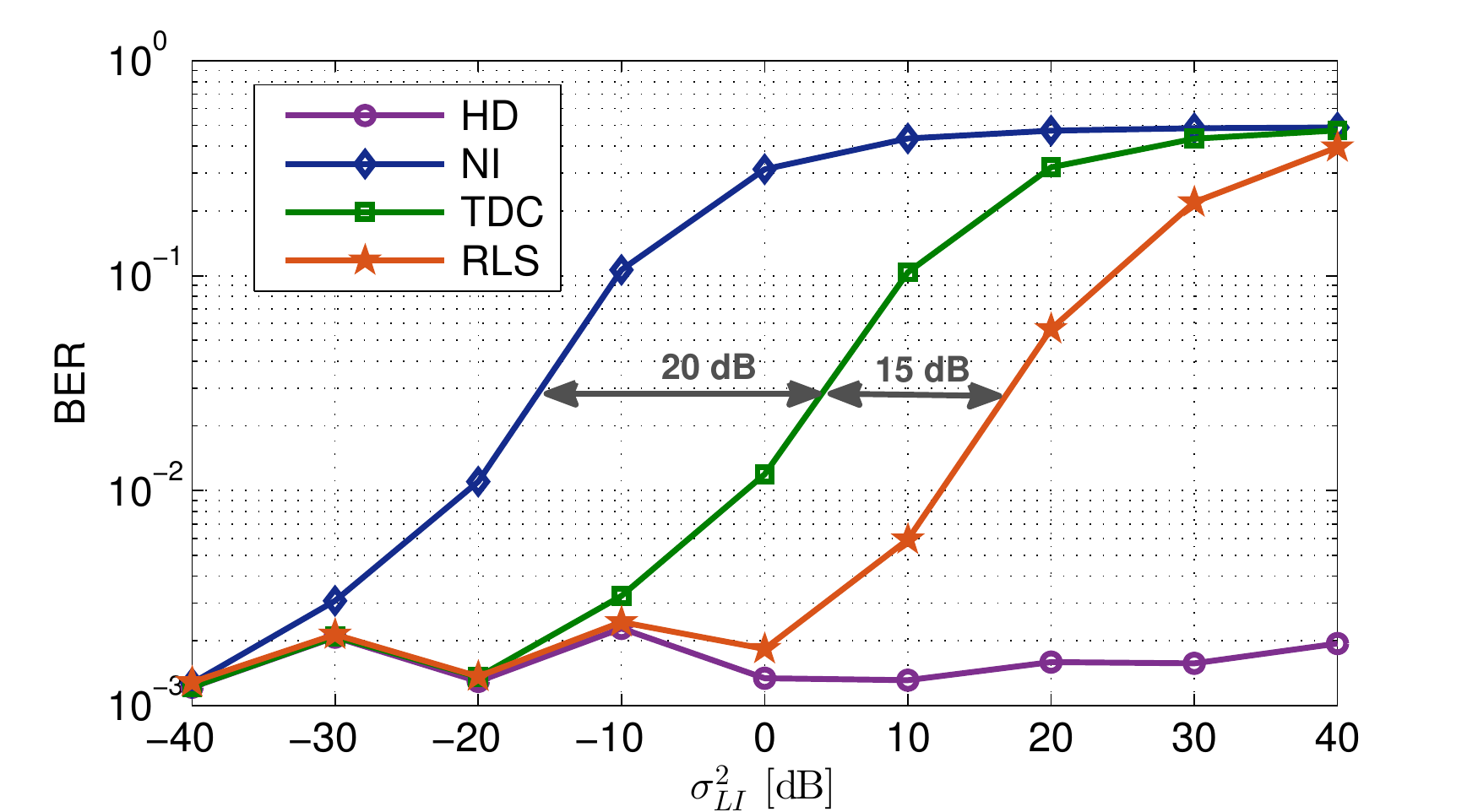}
\end{center}
\caption{BER comparison of the proposed RLS algorithm, TDC and NI, for a 16-QAM OFDM zero-forcing detector.}
\label{fig:BER}
\end{figure} 
   
We now evaluate the SINR and the BER for the parameters considered above and for different values of interference power $\sigma^2_{LI}$. We compare the proposed cancellation technique with the case where there is no filtering, $\mathbf{A_{\star}}=\mathbf{0}$, known as natural isolation (NI), and TDC with a channel estimation error of $-20$ dB, i.e., $\mathbf{\mathcal{E}_{H_{LI}}}[k](n)  \sim \mathcal{CN}(0,\alpha \sigma^2_{LI} \mathbf{I})$ with $\alpha=10^{-2}$, as in section \ref{tdc}. The BER gain against an equivalent half-duplex (HD) system is also assessed. 
Figure~\ref{fig:SINR} shows the $\text{SINR}$ curves, defined as $P_\mathbf{x}/(P_\mathbf{i}+P_\mathbf{n_R})$, where $\mathbf{i}(n)=\mathbf{f}(n)+\mathbf{z}(n)$ is the remaining interference component after cancellation, for the three considered cases, when the RLS algorithm has converged. There, we can observe the large gain provided by the RLS cancellation, which is not affected by erroneous channel estimation as the TDC case.   
The provided SINR gain in Figure \ref{fig:SINR} consequently translates into a BER curve gain, as shown in Figure \ref{fig:BER}. By applying the proposed RLS filter, it is possible to obtain a gain of around 15 dB of self-interference resilience when compared to the case of TDC. Altogether, approximately 35 dB are obtained when taking the NI performance as the base level, which surpasses the 28.6 dB margin obtained in \cite{emilio:adapt}.


\section{Conclusions}
\label{sec:conclusions}

It was shown that the multi-dimensional RLS filter is capable of efficiently estimate the self-interference in an in-band full-duplex decode-and-forward broadband MIMO relay. 
The paper first derived the MIMO RLS filter in the context of full-duplex relaying and then assesses its performance in comparison to other traditional mitigation techniques.
A computationally efficient filtering update rule was defined for this application and it was latter shown that it assures convergence to the optimal MSE estimation. The convergence time of the algorithm was numerically evaluated and shown to be negligible when compared to a practical wireless OFDM transmission block. Finally, the simulation of a full-duplex relay system allowed us to evaluate the interference resilience of the proposed technique, in terms of SINR and BER. In the case of both these metrics, one can found a very significant gain when compared to previously proposed signal processing-domain mitigation techniques.

\bibliographystyle{IEEEtran}

\IEEEtriggeratref{9}

\bibliography{biblio}

\end{document}